# OSHI - Open Source Hybrid IP/SDN networking
# (and its emulation on Mininet and on distributed SDN testbeds)

Stefano Salsano[(1)], Pier Luigi Ventre[(2)], Luca Prete[(2)], Giuseppe Siracusano[(1)], Matteo Gerola[(3)], Elio Salvadori[(3)]
(1) CNIT / Univ. of Rome Tor Vergata - (2) Consortium GARR - (3) CREATE-NET

*Abstract* - The introduction of SDN in IP backbones requires the coexistence of regular IP forwarding and SDN based forwarding. The former is typically applied to best effort Internet traffic, the latter can be used for different types of advanced services (VPNs, Virtual Leased Lines, Traffic Engineering…). In this paper we first introduce the architecture and the services of an "hybrid" IP/SDN networking scenario. Then we describe the design and implementation of an Open Source Hybrid IP/SDN (OSHI) node. It combines Quagga for OSPF routing and Open vSwitch for OpenFlow based switching on Linux. The availability of tools for experimental validation and performance evaluation of SDN solutions is fundamental for the evolution of SDN. We provide a set of open source tools that allow to facilitate the design of hybrid IP/SDN experimental networks, their deployment on Mininet or on distributed SDN research testbeds and their test. Finally, using the provided tools, we evaluate key performance aspects of the proposed solutions. The OSHI development and test environment is available in a VirtualBox VM image that can be downloaded.

## I. Introduction

Software Defined Networking (SDN) [1] is proposed as the new paradigm for networking that may drastically change the way IP networks are run today. Important applications have been found in Data Centers and in corporate/campus scenarios and there have been several proposals considering SDN applicability in wide area IP networks. SDN is based on the separation of the network control and forwarding planes. An external SDN controller can (dynamically) inject rules in SDN capable nodes. According to these rules the SDN nodes perform packet inspection, manipulation and forwarding. They can inspect and modify packet headers at different levels of the protocol stack, from layer 2 to application layer.

Figure 1 shows an example of a single provider domain network interconnected with other providers using BGP. Within the provider network, an intra-domain routing protocol like OSPF is used. The provider offers Internet access to its customers, as well as other transport services (e.g. layer 2 connectivity services or more in general VPNs). Using the terminology borrowed by IP/MPLS networks, the provider network includes a set of Core Routers (CR) and Provider Edge (PE) routers, interconnected by point to point links (POS, GBE, 10GBE…) or by legacy switched LAN (including VLANs). The Customer Edge (CE) routers represent the IP based customer devices connected to the provider. Most often, a provider network integrates IP and MPLS technologies. MPLS creates *tunnels* (LSP – Label Switched Path) among routers. On one hand, this can be used to improve the forwarding of regular IP traffic with: i) traffic engineering, ii) fault protection and iii) avoiding the distribution of the full BGP routing table to intra-domain transit routers. On the other hand, MPLS tunnels are used to offer VPNs and layer 2 connectivity services to customers. In any case, all MPLS implementations are based on a traditional (locked) control plane architecture that does not leave much space for introducing innovation in an open manner.

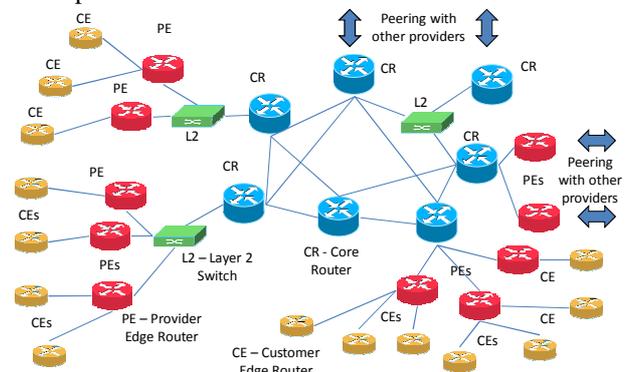

Figure 1. Reference Network scenario

Let us consider the migration of a provider network to SDN. IP core and access routers could be replaced by SDN capable switches, giving the possibility of realizing advanced and innovative services and/or optimizing the provisioning of the existing ones. The migration paths should foresee the coexistence of IP and SDN based services, in an hybrid IP/SDN scenario (resembling the current coexistence of IP and MPLS). In this migration scenario a set of hybrid IP/SDN nodes are capable of acting as plain IP routers (running the legacy IP routing protocols), as well as SDN capable nodes, under the control of SDN controllers. We observe that IP/MPLS control and forwarding plane have been optimized in the years and are capable to operate on large scale carrier networks, while SDN technology has not reached the same maturity level. Therefore, the advantages of introducing SDN technology in a carrier grade IP backbone is not related to performance improvements. Rather we believe that the openness of the SDN approach can ease the development of new services and can foster innovation.

Pure SDN solutions based on L2 switches interconnected with a centralized controller have been shown to work in data-centers and in geographically distributed research networks, such as OFELIA [7] in EU, GENI and Internet2 in US. We argue that an ISP network requires a more sophisticated approach that leverages also on distributed protocols and is interconnected with L2/L3 "standard" networks. As stated in [17], an hybrid SDN model that combines SDN and traditional architectures may "sum their benefits while mitigating their respective challenges". The recent literature presents some preliminary architectures toward hybrid SDN networks (Google B4 WAN [5] can be considered an example).

In [2] the authors presented an open-source Label Switching Router that generates OSPF and LDP packets using Quagga [13] and thus computes the MPLS labels that are installed in the switches using the OpenFlow (OF) protocol. This architecture provides a distributed "standard" control plane, while it uses OF only locally in a node to synchronize the FIBs and to program the data plane. RouteFlow [3] leverages on Quagga to compute the routing that is eventually installed into the hardware switches via the OF protocol. RouteFlow creates a simulated network, copy of the physical one, in the OF controller machine, and exploits distributed protocols, such as OSPF, BGP, IS-IS between the virtual routers of the simulated network.

Compared with these works, our solution considers hybrid IP/SDN nodes also capable of dealing with IP routing, thus achieving easier interoperability with non-OF devices in the core of the network and fault-tolerance based on the regular IP routing. The idea of supporting retro-compatibility and incremental deployment of SDN is not new. According to the OpenFlow specifications [4], two types of switches are supported: OF-only and OF-hybrid which is supporting both OF processing and standard L2/L3 functionalities. Currently, only proprietary hardware switches implement the hybrid approach, with L3 "standard" routing capabilities; in this paper we analyzed and implemented a fully open-source OF-hybrid solution designed to be flexible and scalable which is aimed at facilitating experimentation on hybrid SDN at scale.

The contributions of this paper are multifold: 1) High level design of an hybrid IP/SDN network. 2) Design and implementation of an hybrid IP/SDN node made of open source components, called Open Source Hybrid IP/SDN (OSHI). 3) Design and implementation of an open reference environment to deploy and test the OSHI nodes and related network services. 4) Performance evaluation of some key aspects of the OSHI node and of the overlay topology deployment approach.

The source code of all the components of the OSHI node and of the different tools that have been developed has been published at [9]. In order to ease the initial setup of the solution for other researchers, everything has also been packaged in a ready-to-go virtual machine (available at [9]), with pre-designed example topologies in the order of 30 nodes. To the best of our knowledge, there is no such hybrid IP/SDN node readily available in Open Source.

II. HYBRID IP/SDN NETWORKS

In current IP/MPLS scenario, there is a clear notion of the MPLS tunnels, called LSPs (Label Switched Paths). In a SDN network several types of tunnels or more generically *network paths* can be created, exploiting various fields of different protocols (TCP/UDP, IP, VLANs, Ethernet, MPLS, …). There is not a standard established terminology for such concept, we will refer to these network paths as *SBP*: *SDN Based Paths*. A SBP is a "virtual circuit" which is setup using SDN technology to forward a *packet flow* between two SBP end-points across a set of SDN capable nodes. The notion of packet flow is very broad and it can range from a *micro-flow* i.e. a specific TCP connection between two hosts, to a *macro-flow* e.g. a collection of traffic among different subnets. A flow can be identified looking at headers at different protocol levels.

We address the definition of the Hybrid IP/SDN network by considering: i) mechanisms for coexistence of regular IP traffic and SBPs; ii) the set of services that can be offered using the SBPs; iii) ingress classification and tunneling mechanisms.

Let us consider the coexistence of regular IP traffic and SDN based paths on the links between Hybrid IP/SDN nodes. A SDN approach offers a great flexibility and can classify packets with a "cross-layer" approach, by considering packet headers at different protocol levels (MPLS, VLANs, Q-in-Q, Mac-in-Mac and so on). Therefore it is possible to specify a set of conditions regarding flows that have to be handled at IP level and the ones to be handled using SDN. These conditions can be in the form of white lists / black lists and can change dynamically, interface by interface. This flexibility may turn into high complexity, therefore the risk of misconfigurations and routing errors should be properly taken into account (see [6]). In the end, different coexistence and tunneling mechanisms will operate in a Hybrid IP/SDN network and MPLS encapsulation could be among the preferred options. Coming to our Open Source demonstration we took into account the capability of available tools in terms of tunnels handling and of compliance to OpenFlow protocol releases. Therefore we decided to use VLAN tags as IP/SDN coexistence mechanisms and designed two solutions: i) the IP traffic travels in the network with a specific VLAN tag while SBPs use other VLAN tags (Tagged coexistence); ii) the IP traffic travels "untagged" and SBPs use VLAN tags (Untagged coexistence). The proposed VLAN based mechanism can interwork with existing legacy VLANs, as long as a set of tags is reserved for our use on each VLAN (and this is needed only if the link between two routers already runs on a VLAN). The VLAN mechanism used in our implementation can be replaced by other more scalable tunneling mechanisms like MPLS without changes in the proposed architecture and we are now working on it.

Let us now consider the services and features that can be offered by the Hybrid IP/SDN network. As of writing, we designed and implemented an "Ethernet Virtual Leased Line" (VLL). This service guarantees to the service end-points to be directly interconnected as if they were in the same Ethernet LAN. Our implementation offers services at an edge router, the end-point can be a physical port of the edge router or a logical port (i.e. a specific VLAN). Two arbitrary end-points in edge routers can be bridged by the offered VLL service. The interconnection is realized in our Hybrid IP/SDN network with a SBP using VLAN tags. As further work we are now considering the offering of an Ethernet Virtual Switch, in which several end-points can be transparently bridged into a virtual switch offered by the Hybrid IP/SDN network.

Let us finally consider the ingress classification and tunneling functionality. The ingress edge router will need to classify incoming traffic as belonging to IP best effort or as traffic to be encapsulated traffic in the SBPs considering the selected tunneling approach. The egress edge router will extract the traffic from the SBPs and forward it to the appropriate destination. The ingress classification mechanisms in our testbed takes into account the requirement of VLL service. We can classify untagged traffic entering in a port of an edge router as regular IP traffic or as belonging to the end-point of a VLL depending on the input port; for VLAN tagged traffic, each VLAN tag can be linked to a VLL end point and one of the VLAN tags can be assigned to regular IP traffic.

## III. OSHI NODE ARCHITECTURE

We designed our Open Source Hybrid IP/SDN (OSHI) node combining a SDN Capable Switch (SCS), an IP forwarding engine and an IP routing daemon. The SDN Capable Switch is connected to the set of physical network interfaces belonging to the integrated IP/SDN network, while the IP forwarding engine is connected to a set of virtual ports of the SCS, as shown in Figure 2. In our OSHI node, the SCS component is implemented using Open vSwitch (OVS) [12], the IP forwarding engine is the Linux kernel IP networking and Quagga [13] acts as the routing daemon.

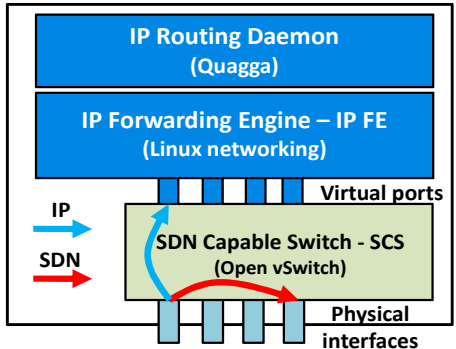

Figure 2. OSHI Hybrid IP/SDN node architecture (data plane)

The internal virtual ports that interconnect the SCS with the IP forwarding engine are realized using the "Virtual Port" feature offered by Open vSwitch. Each virtual port is connected to a physical port of the IP/SDN network, so that the IP routing engine can reason in term of the virtual ports, ignoring the physical ones. The SCS differentiates among regular IP packets and packets to be processed by SCS. By default, it forwards the regular IP packets from the physical ports to the virtual ports, so that they can be processed by the IP forwarding engine, controlled by the IP routing daemon. This approach avoids the need of translating the IP routing table into SDN rules to be pushed in the SCS table. A small performance degradation is introduced because a packet to be forwarded at IP level will cross the SCS switch twice. It is possible to extend our implementation to consider mirroring of the IP routing table into the SCS table, but this is left for further work.

An initial configuration of the SCS tables is needed to connect the physical interfaces and the virtual interfaces, to support the SCS-to-SDN-controller communication, or for specific SDN procedures (for example to perform Layer 2 topology discovery in the SDN controller). A local management entity in the OSHI node takes care of these tasks. In our setup, it is possible to use an "in-band" approach for SCS-to-SDN-controller, i.e. using the regular IP routing/forwarding, avoiding the need of setting up a separate out-of-band network. Further details and the block diagram of the control plane architecture of OSHI nodes are reported in [10].

As for the implementation of coexistence mechanisms between IP and SDN traffic, services and ingress classification we leveraged on the multiple tables functionality introduced with OpenFlow v1.1 and recently implemented in Open vSwitch. The details are also reported in [10].

The VLL service is implemented with a SBP that switches VLAN tags between two end-points (in both directions). The creation of the SBP is performed using a python script called VLLPusher. It uses the Topology REST API of the Floodlight controller in order to retrieve the route that interconnects the VLL end-points. It allocates the VLAN tags and then uses the Static Flow Pusher REST API to set the rules for packet forwarding and VLAN tag switching.

The implementation of the ingress classification is realized within the SCS of Access OSHIs. By configuring rules in the SCS it is possible to map the traffic on an ingress physical port as follows: i) untagged traffic to a virtual port (for regular IP); ii) untagged traffic to a SBP (for a VLL end-point); iii) VLAN tagged traffic to a virtual port (for regular IP); iv) VLAN tagged traffic to a SBP.

## IV. OSHI EMULATION TOOLS

We realized our OSHI node for three target environments: Virtual Box, Mininet emulator and the OFELIA testbed [7] (using XEN virtualization). We used the Virtual Box deployment to emulate small setups with two or three OSHI nodes (details in [9]). Hereafter we focus on the experiments on Mininet and OFELIA, based on the emulation workflow shown in Figure 3.

### A. Topology and Service generation

We developed a JavaScript web GUI (TopoDesigner), which allows to design a network topology and to configure the services (see Figure 9). It exports the created topology in JSON format. It is also possible to synthetically generate a topology using Networkx [8], a Python package for the creation/manipulation of complex networks. A set of python scripts (Topology Deployer) parses the topology file and deploys the experiment over Mininet or OFELIA.

## B. Mininet deployment

As for the Mininet deployment, we extended the functionality of the emulator. By default, Mininet provides only hosts and switches (the latter ones with either user space or kernel space implementation). We have introduced an extended host capable to run as a router and managed to run Quagga and OSPFD daemons on it. Then we have added to it the OVS functionality in order to realize the OSHI node. The details on the architecture for the deployment on Mininet can be found in [10]. The Mininet Deployer is able to automate all the aspects of an experiment. This includes the automatic configuration of IP addresses and of dynamic routing (OSPF daemons) in all nodes, therefore it relieves the experimenter from a huge configuration effort.

## C. OFELIA deployment

As for the OFELIA deployment, we run our experiments on the CREATE-NET testbed based on the OCF (OFELIA Control Framework) developed in the context of the OFELIA project [7]. The testbed is composed by a set of 8 OpenFlow capable switches and 3 Virtualization Servers that can host experimental Virtual Machines controlled by the testbed experimenters. Using SDN mechanisms (and in particular the Flowvisor network virtualization tool [14]) the testbed resources can be "sliced" among different experiments and each experiment can be controlled by a different OpenFlow controller. Our deployer can actually operate on any OCF compliant testbed (several testbed "islands" based on OCF are currently available as a result of the OFELIA project).

Different mechanisms have been used to automate and facilitate both the setup and the configuration processes. A management server coordinates all operations, communicating with the rest of the experiment machines through the testbed management network and using Distributed Shell (DSH) for distributing and executing remote scripts. Through the setup procedure, the needed scripts are copied on the experiment VMs and run locally. A more detailed overview of the scripts and a typical use-case can be found at [9].

As shown in Figure 3, the OFELIA Deployer python script automatically produces the configuration scripts for emulating a given topology, composed of access and core OSHI nodes (PE and CR) and of Customer Edge routers (CE). This includes the automatic configuration of IP addresses and of dynamic routing (OSPF daemons) in all nodes, therefore it relieves the experimenter from a huge configuration effort.

Each OSHI node and each CE is mapped into a different VM running in one of the Virtualization Server of the testbed. The Topology to testbed mapping file contains the association of OSHI nodes and CEs to testbed VMs. An overlay of Ethernet over UDP tunnels is created among the VMs to emulate the network links among OSHI nodes and between OSHI nodes and CEs.

In our first design we created the tunnels using the user space OpenVPN tool (with no encryption). The performance was poor, as performing encapsulation in user space is very CPU intensive. Therefore we considered a second design approach that uses VXLAN [16] tunnels provided by Open vSwitch (OVS). As explained in [15], OVS implements VXLAN tunnels in kernel space, and this allows to dramatically improve performance with respect to OpenVPN. The design of the VXLAN tunneling solution for OSHI over an OFELIA testbed is reported in Figure 4. We only use VXLAN as a point-to-point tunneling mechanism (the VXLAN VNI identifies a single link between two nodes) and we do not need underlying IP multicast support as in the full VXLAN model. The SDN capable OVS is also able to perform encapsulation and decapsulation of VXLAN tunnels, each tunnel corresponds to a port in the switch. The VXLAN tunnel ports in Figure 4 are conceptually equivalent to physical ports shown in Figure 2.

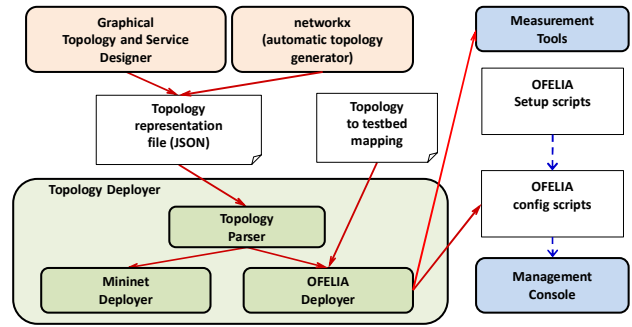

Figure 3. OSHI emulation deployment workflow

In order to automate as much as possible the process of running the experiments and collecting the performance data we have developed an object oriented multithreaded python library called OSHI-MT (Measurement Tools). Using the library we can remotely run the traffic generators (iperf) and gather load information (CPU utilization) on all nodes (VMs). As for the load monitoring, taking CPU measurements from within the VMs (e.g. using the *top* tool) does not provide reliable measurements. The correct information about the resource usage of each single VM can be gathered with the *xentop* tool, which must be run as root in the hosting XEN server. Therefore we have developed a python module that collects CPU load information for each VM of our interest in the XEN host using *xentop* and formats it in a JSON text file. OSHI-MT retrieves the JSON file from the python module with a simple message exchange on a TCP socket.

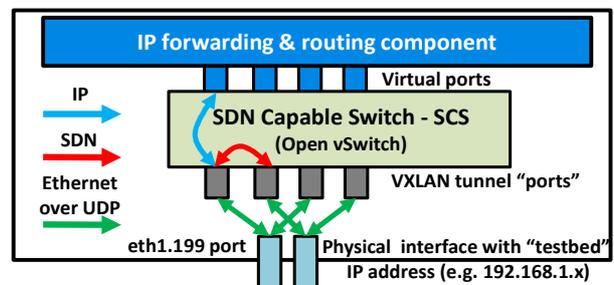

Figure 4. Implementing VXLAN tunnels using Open vSwitch (OVS)

## V. PERFORMANCE EVALUATION

In this section we analyze some performance aspects with 4 experiments.

i) IP forwarding performance using OSHI, with reference to the data plane architecture shown in Figure 2
ii) Comparison of OpenVPN and VXLAN tunneling for experiments over distributed SDN testbeds
iii) IP forwarding performance (plain and OSHI) and SBP forwarding in an overlay topology using tunneling mechanisms over distributed SDN testbeds
iv) IP forwarding and SBP performance over Mininet

In the experiments i, ii and iii we use the iperf tool as traffic source/sink in the CE routers and generate UDP packet flows from 500 to 2500 packet/s (datagram size is set at 1000 Byte). We evaluate the CPU load in the PE routers with our xentop based measurement tool. We executes periodic polling and gather the CPU load of the monitored VMs. In each run we collect 20 CPU load samples with polling interval in the order of two seconds, the first 10 samples are discarded and the last 10 are averaged to get a single CPU load value. Then we evaluate the mean (AVG) and standard deviation (DEV) over 20 runs.

### A. First performance assessment of OSHI

We consider the design in Figure 2 and compare the OSHI IP forwarding (each packet crosses the Open vSwitch two times) with ROUTER IP forwarding (the Open vSwitch is removed and the OSHI node interfaces are directly connected to IP forwarding engine). We refer to this scenario as "plain VLAN" as no tunneling mechanism is used. This experiment is not automatically deployed using the topology designer and deployer, and we setup a limited topology with two End User Hosts and two OSHI nodes. In this case it is not possible to deploy the VLL service and only plain IP router and OSHI IP have been compared. In the experiment results (see [10] for details) we can appreciate a CPU load penalty for OSHI IP forwarding with respect to plain IP forwarding in the order of 10%-20%. Apparently, the CPU load penalty is decreasing in relative terms at higher CPU load, but this is subject to further evaluation in future experiments. The theoretical CPU saturation rate for plain IP router is in the order of 14000 p/s. Adding OSHI-IP forwarding reduces the theoretical CPU saturation rate to something in the order of 12500 p/s.

### B. Performance comparison of OpenVPN and VXLAN tunneling

In this experiment we have deployed the topology represented in Figure 6 over the physical testbed topology in Figure 5 (3 OpenFlow switches and 3 virtualization servers). We want to evaluate the processing overhead introduced by the tunneling mechanisms (OpenVPN and VXLAN) used to deploy the overlay experimental topologies over distributed SDN testbeds.

In Figure 7 we compare the CPU load for OSHI IP solution in the OpenVPN, VXLAN and plain VLAN scenarios. It can be appreciated that VXLAN tunneling adds a reasonably low processing overhead, while OpenVPN tunneling would dramatically reduce the forwarding capability of an OSHI node in the testbeds. The theoretical CPU saturation rate for OpenVPN tunneling is in the order of 3500 p/s, which is 4 times lower than in the plain VLAN case. The theoretical CPU saturation rate for VXLAN tunneling is only ~8% lower than the plain VLAN case, showing that VXLAN is an efficient mechanism to deploy overlay topologies.

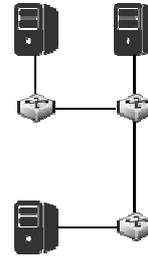
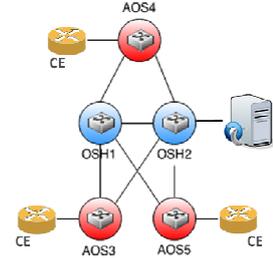

Figure 5. Physical network    Figure 6. Overlay network

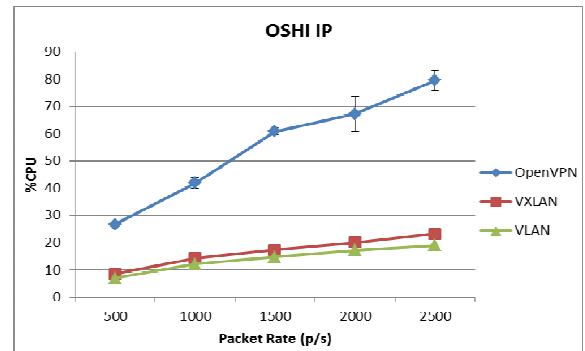

Figure 7. CPU Load for different tunneling mechanisms.

### C. Performance analysis on distributed SDN testbed

In this set of experiment we evaluate the processing load of different forwarding solution over the distributed SDN testbeds considering the topology shown in Figure 6. For the OSHI solution, we consider IP forwarding (OSHI IP) and SBT forwarding (OSHI VLL). Then we consider plain IP forwarding as a reference (ROUTER IP).

We executed the performance tests of OSHI IP, OSHI VLL and ROUTER IP using the VXLAN tunneling solution. Figure 8 provides reports the experiment results. As shown in Figure 4, in this case the plain IP forwarding (ROUTER IP) has to go through Open vSwitch which handles the VXLAN tunneling in any case, therefore as expected it has no advantage with respect to OSHI IP. The OSHI VLL solution is the less CPU intensive and its theoretical CPU saturation rate is in the order of 13000 p/s. The OSHI IP solution increases CPU load of less than 10%, and its theoretical CPU saturation rate is in the order of 12000 p/s.

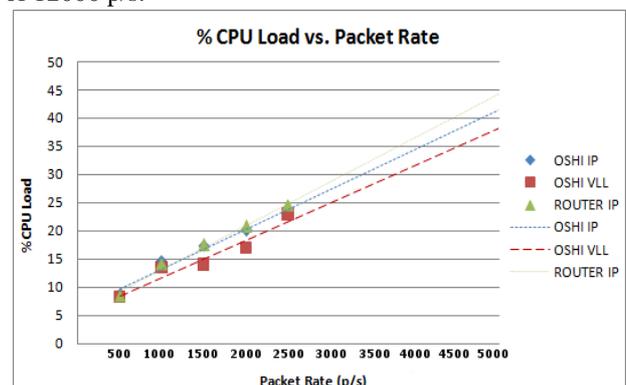

Figure 8. CPU load with VXLAN tunneling.

## D. Experimentation with Mininet

The fourth experiment has been run over the Mininet emulator. We have deployed the topology shown in Figure 9, which reports the image from the GUI of the TopoDesigner. On this topology, we evaluate the TCP throughput between two End User Hosts comparing the OSHI IP solution with the OSHI VLL service. We argue that the throughput will be limited by the sum of CPU processing load on all nodes (the Mininet emulation is run on a single machine that emulates the whole network). Therefore we expect that the throughput using the OSHI VLL service will be higher than the throughput using the OSHI IP routing. This is confirmed by the results shown in Table 1, which shows a 26% increase of the throughput using the VLL with respect to using IP routing.

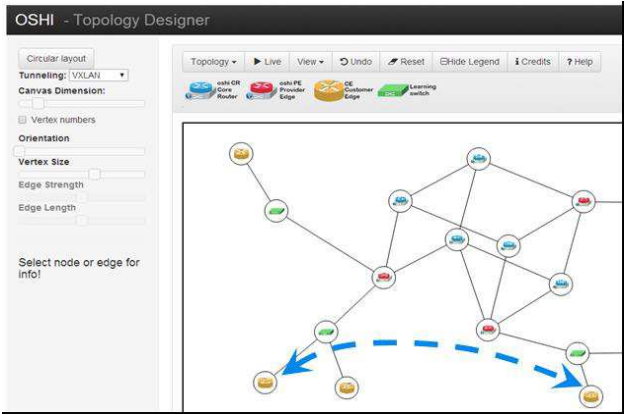

Figure 9. Mininet topology and deployed VLL

|  | VLL (Mb/s) | IP (Mb/s) |
|---|---|---|
| AVG | 1555 | 1150 |
| STD DEV | 21,8 | 20 |

Table 1 TCP max throughput for OSHI IP and VLL

## VI. CONCLUSIONS

In this paper we presented a novel architecture and implementation for an Hybrid IP/SDN (OSHI) node. Developed according to an open-source model, OSHI is intended to enable researchers to investigate novel applications in the emerging hybrid IP/SDN carrier networks scenario. Results out of performance tests executed both in single-host emulators (Mininet) and in distributed SDN testbeds (OFELIA) demonstrate that OSHI is suitable for large scale experimentation settings.

## ACKNOWLEDGMENTS

This work was partly funded by the EU in the context of the DREAMER project, one of the beneficiary projects of the GÉANT Open Call research initiative.